\documentstyle[11pt,paspconf]{article}

\input psfig

\def\gtorder{\mathrel{\raise.3ex\hbox{$>$}\mkern-14mu
             \lower0.6ex\hbox{$\sim$}}}
\def\ltorder{\mathrel{\raise.3ex\hbox{$<$}\mkern-14mu
             \lower0.6ex\hbox{$\sim$}}}

\begin{document}

\title{The Evolution of Gravitational Lens Galaxies }

\author{C.S. Kochanek\altaffilmark{1}, E.E.~Falco\altaffilmark{1},
  C.D.~Impey\altaffilmark{2}, J.~Leh\'ar\altaffilmark{1}, 
  B.A.~McLeod\altaffilmark{1}, H.-W. Rix\altaffilmark{3},
  C.R. Keeton\altaffilmark{2}, J.A. Mu\~noz\altaffilmark{1} 
  \& C.Y.~Peng\altaffilmark{2}}

\altaffiltext{1}{Harvard-Smithsonian Center for Astrophysics}
\altaffiltext{2}{Steward Observatory, University of Arizona} 
\altaffiltext{3}{Max-Planck-Institut fuer Astronomie, Heidelberg}

\begin{abstract}
Most gravitational lens galaxies are early-type galaxies in relatively
low density environments.  We show that they lie on the same fundamental
plane as early-type galaxies in both local and distant rich clusters.
Their surface brightness evolution requires a typical star formation 
epoch of $z_f\simeq2$-$3$, almost indistinguishable from that
of rich cluster galaxies at comparable redshifts.  The restricted galaxy
type range of the lenses means that photometric redshifts work 
well even with only 1--3 filter photometry.  We make preliminary 
measurements of the mass and luminosity functions of the lens galaxies,
and find they are consistent with the standard model used for deriving
cosmological limits using lens statistics.
As expected for a mass-weighted sample, they are more massive and more
luminous than the overall early-type galaxy population.
\end{abstract}


\section{Introduction}

Gravitational lens galaxies are a unique sample, because they are the
only galaxies selected based on mass rather than luminosity.  As such,
the average properties of the lens galaxies are {\it identical} to the
mass-weighted average properties of {\it all} galaxies at a given 
redshift.  At
intermediate redshifts ($0 < z_l < 1$), they are also the largest sample
of galaxies with known masses outside the cores of rich clusters.
Historically, the interpretation of the mass measurements has been 
hampered by a lack of accurate measurements of the {\it light}. 
Eliminating this rather peculiar problem for an astronomical sample
is a primary goal of the CASTLES (CfA/Arizona Space Telescope Lens Survey) 
project (see Falco et al. in these proceedings).  

If we combine accurate measurements of the mass and the light with the wide
range of lens redshifts, then we have an excellent tool for studying the 
evolution of galaxy mass-to-light (M/L) ratios with redshift.  Our 
method is similar to that introduced by van Dokkum \& Franx (1996)
and used by Kelson et al. (1997, 1999), van Dokkum et al. (1998, 1999),
Pahre et al. (1999ab), and Jorgensen et al. (1999) to measure the 
evolution of early-type galaxies in
rich clusters over the same redshift range.  The lens galaxies are
in far lower density environments than those in rich clusters -- a ``field'' rather than a 
cluster population -- and theoretical models (e.g. Kauffmann 1996, Kauffmann \& 
Charlot 1998) predict that they should have significantly younger
stellar populations than galaxies in rich clusters.  Studies of the
colors and luminosity-effective radius correlations (e.g. Schade et al. 1996, 1999,
Ziegler et al. 1999), and
one small fundamental plane sample (Treu et al. 1999) all suggest that there is 
little difference between the field and cluster populations at $z\sim 0.5$.
With the lens galaxies we can directly measure the surface brightness
evolution to $z\simeq1$. 

In this review we summarize our recent results on galaxy evolution
(Kochanek et al. 1999) and illustrate new results on the mass and
luminosity functions of the lens galaxies. In \S2 we introduce
the fundamental plane (FP) of lens galaxies, which we use to 
determine the star formation epoch in \S3 and as a photometric redshift 
estimator in \S4. In \S5 we determine the mass and luminosity function 
of the lenses as compared to local galaxies and simple statistical models,
and in \S6 we review our goals for the future.  

\begin{figure}[hp]
\centerline{\psfig{figure=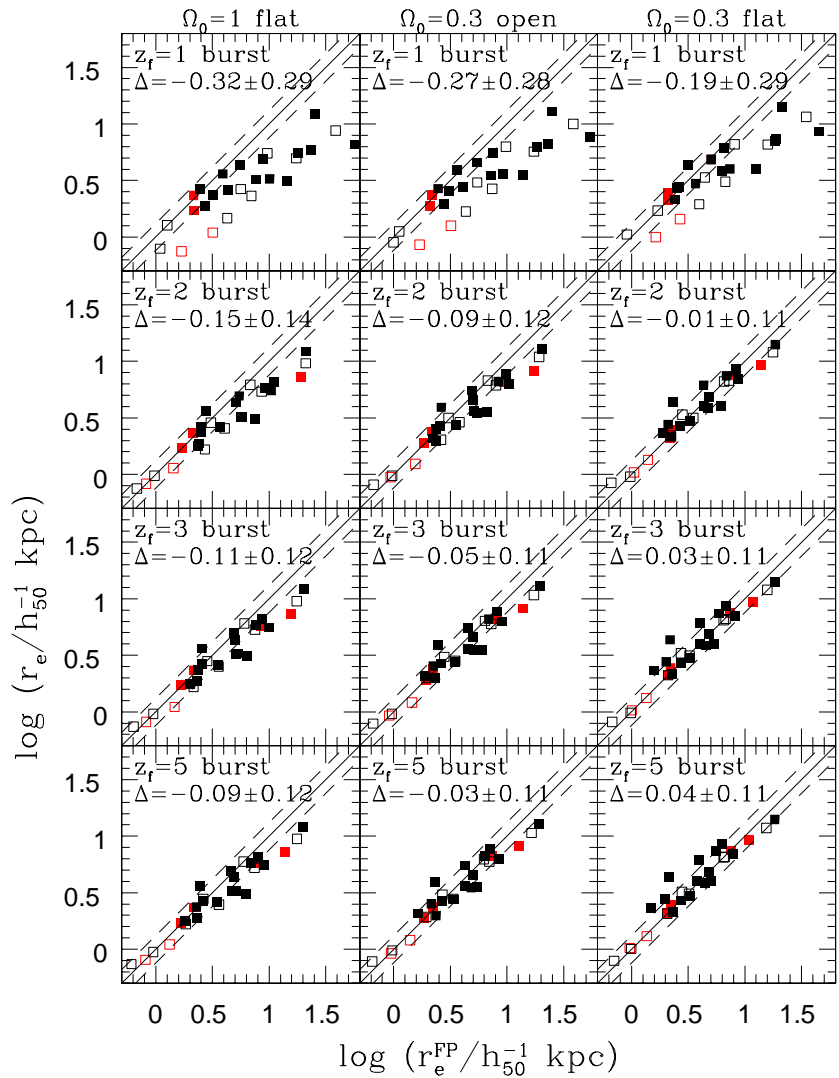,width=4.5in}}
\caption{The FP of lens galaxies transformed to zero redshift.  The cosmologies
  (from left to right) are the $\Omega_0=1.0$ flat, $\Omega_0=0.3$ open and
  $\Omega_0=0.3$ flat models.  An instantaneous burst star formation history
  is used with star formation redshifts (from top to bottom) of $z_f=1$, $2$,
  $3$ and $5$.  The filled squares are for the lenses with known redshifts and
  for the open squares we have used photometric redshifts (see \S4).  The solid 
  line marks the FP of the local comparison sample;
  90\% of the galaxies in the local JFK sample lie between the dashed lines.
   The mean residual ($\Delta = \langle \log (r_e/ r_e^{FP})\rangle$) and its dispersion
   are shown in the upper left corner of each panel.  These are calculated using only
   the systems with known lens redshifts. 
  }
\end{figure}

\section{The Fundamental Plane And Galaxy Evolution}

The fundamental plane (FP) is a tight correlation between
the central velocity dispersion $\sigma_c$, intermediate axis effective
radius $r_e$, and central surface brightness $\mu_e$ discovered by 
Djorgovski \& Davis (1987) and Dressler et al. (1987).  The FP is a
consequence of virial equilibrium, the nearly self-similar photometric
and kinematic properties of early-type galaxies, and systematic rather
than random variations in the metallicity and stellar population ages
with mass or luminosity.  As a tool for studying galaxy evolution, its
key feature is that given any two of the three variables one can
predict the remaining one.  Traditionally this is done by comparing
the directly measured effective radius $r_e$ to that predicted by
the FP, 
\begin{equation}
\log(r_e^{FP}/h_{50}^{-1}\hbox{kpc}) = 1.24\log (\sigma_c/\hbox{km s}^{-1})
  + 0.33 \mu_e(0) - 8.66
\end{equation}
using Jorgensen, Franx \& Kjaergaard (1993, 1995ab, 1996, collectively
referred to as JFK) renormalized to the closest HST 
filter (F606W).  The only (rapidly?) evolving quantity appearing in the FP is the
surface brightness, $\mu_e(z)$, through the time variation in the average
luminosity of a stellar population.  Hence, the standard FP is normalized
using the surface brightness today, $\mu_e(z=0)$.  

The key to studying galaxy evolution is the strong variation with redshift of $\mu_e(z)$
as compared to the invariance of $r_e$ and $\sigma_c$.  If we observe a 
galaxy in a series of filters $j$, we can measure the surface brightness $\mu_j$, 
which evolves as
\begin{equation}
   \mu_j(z) = \mu_j(0) + 10\log(1+z) + e_j(z) + k_j(z)
\end{equation}
where $\mu_j(0)$ is the surface brightness at $z=0$, the first redshift
term is the $(1+z)^4$ cosmological dimming, $e_j(z)$ is the evolution correction
for filter $j$, and $k_j(z)$ is the K-correction from the rest frame to the
observed frame.  If we can estimate the ``final'' surface brightness, $\mu_j(0)$,
then we can directly measure the evolutionary terms.
By simply reordering the standard form of the FP, we have
\begin{equation}
\mu_e^{FP}(0)=-3.76 \log (\sigma_c/\hbox{km s}^{-1})
 +3.03 \log(r_e/h_{50}^{-1}\hbox{kpc}) +26.26,
\end{equation}
so if we measure the effective radius and velocity dispersion of an early-type
galaxy at any redshift, {\it we can predict the surface brightness it will
have today at $z=0$.}  Thus, for any early-type galaxy we have that
\begin{equation} 
     e_j(z) + k_j(z) = \mu_j(z) - \left[\mu_j^{FP}(0)+10\log(1+z)\right].
\end{equation}
We can determine $e_j(z)$ alone by measuring colors and
interpolating to a fixed rest wavelength (van Dokkum \& Franx 1996).  The uncertainty for any
one galaxy is not just the measurement uncertainty, as we must include the
significant surface brightness scatter of the local FP ($0.23$~mag/arcsec$^2$
in JFK).

Our procedures differ somewhat from the cluster studies because we must consider
individual lens galaxies rather than samples of cluster galaxies at a common 
redshift.
The cluster studies measured the properties of enough galaxies in each
cluster (6 on average) to use the standard FP (eqn. 1), measure the change in
the zero-point (the 8.66) with redshift and then interpret it as evolution in
the population averaged surface brightness or mass-to-light ratio.  Since we
cannot construct directly the FP of lens galaxies at any redshift we are driven
to our alternate, galaxy by galaxy formulation.  We believe, however, that 
our formulation of the analysis is the more physical.  The existence of a thin
FP at higher redshifts is observational evidence that early-type galaxies of
similar mass have small age differences, but it is not required for the analysis. 
 
For most lenses we do not have velocity dispersion estimates.  We instead
infer the velocity dispersion from the image separation $\Delta\theta$ assuming
a singular isothermal sphere (SIS) mass distribution.  Dynamical models of galaxies
in SIS halos (Kochanek 1994) demonstrated that the dark matter dispersion, which
we measure with $\Delta\theta$, is coincidentally almost identical to the central
stellar dispersion.  Adopting a normalization scale of $\sigma_{c*}=225$~km~s$^{-1}$
for an $L_*$ early-type galaxy (as required by models of the distribution of lensed
image separations, Kochanek 1996, Falco et al. 1998), our velocity dispersion
estimate becomes
\begin{equation}
   \sigma_c = (225/f)(\Delta\theta/2\farcs91)^{1/2}(D_{OS}/D_{LS})^{1/2} \hbox{km s}^{-1},
\end{equation}
where $f=\sigma_{dark}/\sigma_*\simeq 1.0\pm0.1$ is the normalizing factor between
the velocity dispersion of the dark matter and the stars.  We can also use the FP
to estimate that its value is $f=1.06\pm0.07$, providing further confirmation of the earlier
dynamical and lensing results.  We explored constant M/L dynamical models as 
well, but they generally fit the data poorly.
  
Our final current consists of 29 lenses, 12 of which are missing lens redshifts and 6
of which are missing source redshifts. Missing redshifts are the bane of lensing
studies, and we discuss photometric redshift estimates for the lens galaxies in \S4. 
The estimates of galaxy evolution and the photometric redshift estimates for the
lens galaxies are insensitive to our assumptions about unmeasured source redshifts.
We also analyzed a comparison sample of 54 galaxies from the cluster FP studies, so 
that we could compare the evolution of the two populations directly.  For theoretical
models we used the GISSEL96 versions of the Bruzual \& Charlot (1993) spectral 
evolution models assuming $H_0=65$~km~s$^{-1}$~Mpc$^{-1}$.  For simplicity we have
compared the data to solar metallicity models with a single burst of star formation at redshift $z_f$.
Models with extended bursts are also consistent with the data but require earlier
star formation epochs.  

\begin{figure}[t]
\centerline{\psfig{figure=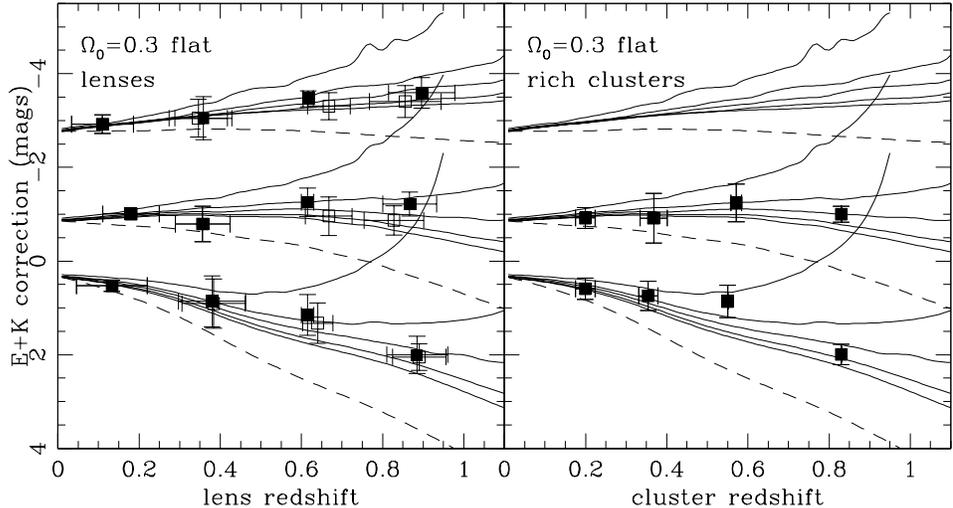,width=5.0in}}
\caption{Evolution and K-corrections for the lens (left) and cluster (right) samples 
in the V=F555W (bottom), I=F814W (middle) and H=F160W (top) bands as a function of redshift 
for a flat $\Omega_0=0.3$ 
cosmological model.  The zero-redshift colors were left in to separate the curves.  
The points are averages in redshift bins with edges at $z=0.25$, $0.50$ and $0.75$.
The error bar on the E+K correction is the standard deviation of the galaxies in
the bin, not the uncertainty in the mean.  For the lenses, the filled points are the 
averages using only the lenses with known redshifts, while the open points
include all lenses.  The dashed curves are the no evolution models for
each filter, and the solid curves are the instantaneous burst models
with star formation redshifts (from bottom to top) of $z_f=10$, $3$,
$2$, $1.5$ and $1.0$ respectively.  }
\end{figure}

\begin{figure}[ht]
\centerline{\psfig{figure=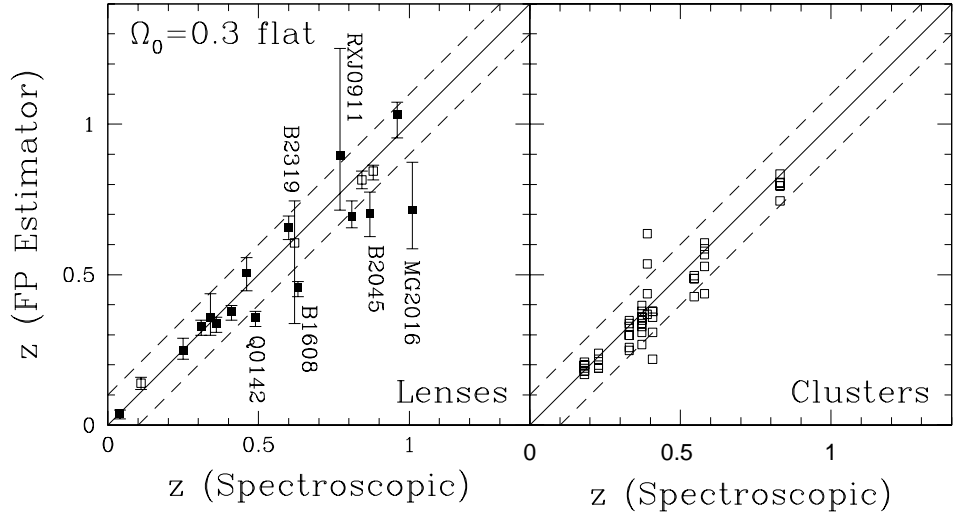,width=5.0in}}
\caption{The FP redshift estimates compared to the spectroscopic redshifts.
  The left panel shows the results for the lens galaxies, and the right for
  the cluster galaxies.  For the lens galaxies, solid (open) points are used for
  lenses where the source redshift is known (unknown).  The dashed lines are
  offset by $|\Delta z|=0.1$ to illustrate the desired accuracy.  We have only
  H-band data for the labeled lenses with large redshift uncertainties (RXJ~0911+0551,
  MG~2016+112, B~2045+265 and B~2319+052). }
\end{figure}

\section{The Star Formation Epoch of Early-Type Galaxies}

The lens galaxies at any given redshift are a mass-weighted sample of {\it all}
galaxies at that redshift.  This means that the mean star formation epoch of 
the lens galaxies is closely related to the mean, mass-weighted star formation 
epoch of all galaxies.  Hence, the star formation epoch of the lens galaxies 
is a far more fundamental measurement than the star formation epoch of
galaxies in rich clusters.  While we consider only early-type lenses here,
they are the vast majority of all lenses.  Moreover, even the late-type 
lenses tend to be fairly red and show few signs of active star formation.
The absence of the numerous, luminous, blue, low-mass star forming galaxies
in the lens sample confirms that these galaxies make a negligible
contribution to the total mass in stars. 

We can show that the lens galaxies lie on a coherent FP by evolving their properties 
forward in time and placing them on the present day FP.  The transformations depend
on the star formation epoch, $z_f$, and the cosmological model ($\Omega_0$ and
$\lambda_0$).  Figure 1 shows the results.  The mean logarithmic residual, 
$\Delta=\langle \log r_e/r_e^{FP}\rangle$, is computed only for the lenses with
known redshifts.   As van Dokkum \& Franx (1996) first noted, it is difficult to reconcile the 
high $\Omega_0$ models with a reasonable star formation epoch ($ z_f \ltorder 10$).
For the low $\Omega_0$ models, a coherent FP matching the local observations is
possible for formation epochs $z_f \gtorder 2$.  The scatter about the FP is 
comparable to that for the local FP, so the lens galaxies cannot have an enormous
range of formation epochs.  

Since we are now confident that the lenses lie on the FP, Figure 2 shows the evolution
of the E+K corrections with redshift for both the lens and cluster samples in the
$\Omega_0=0.3$ flat cosmological model.  The results for an $\Omega_0=0.3$ open model
are similar.  The dominant term depends on the wavelength.
In the V band, the K-correction dominates and the galaxies rapidly fade, in the I band,
the E and K corrections balance, and in the H-band, the evolution dominates and the
galaxies steadily become brighter.  At all bands, the galaxies become steadily brighter
than predicted by a no evolution model.  

It is difficult to distinguish the evolution of the
lens and cluster galaxies, and both samples favor star formation epochs of $z_f\simeq2$-$3$.
This strongly contradicts the predictions of Kauffmann \& Charlot (1998), where most
of the field early-type galaxies form their stars at $z_f \ltorder 1$.  Sommerville
\& Primack (1998) have argued more generally that the earlier semi-analytic models
systematically underestimated the epoch of star formation through their choice of
star formation mechanisms.  In their defense, Kauffmann \& Charlot (1998) would argue
that morphologically selected early-type galaxies look old at all redshifts because
the mergers required to produce the early-type morphologies do not occur when the 
stellar populations are young.  This requires rapid number evolution in the early-type
galaxy population to $z=1$, which is probably inconsistent with observations (Lilly et al. 1995, 
Schade et al. 1999).  Moreover, the lens galaxies were selected based on mass, not morphology, 
and while we excluded a few late-type galaxies, there is no significant population of blue, 
star-forming lenses which we have dropped from the analysis. 
  
\section{Photometric Redshifts}

Redshift incompleteness is the bane of many astrophysical applications of gravitational
lenses, even though there is no technical barrier to measuring most of the missing redshifts in the
age of the 8~meter telescope.  This is unfortunate, because you obtain more cosmological information 
with the measurement of a lens redshift than any other single object.  For example, the
distribution of lens redshifts is a powerful cosmological test (Kochanek 1992), but it requires 
high redshift completeness because it is very difficult to make a statistical model for 
incomplete redshift samples.   The difficulty in measuring lens redshifts is further evidence that 
the stellar populations of lens galaxies are old. If the typical lens had any ongoing star formation, 
the redshift measurements would be far easier due 
emission lines usually associated with star formation. 

Photometric redshifts should work well for lens galaxies because of the restricted
galaxy type distribution (mostly early-type galaxies, and no funny dwarf galaxies) and because
we also know the galaxy masses.  We have developed a redshift estimation method
based on the FP.  Placing a galaxy on the FP can accurately estimate a galaxy
redshift with {\it no} color information if the lens has $z_l \ltorder 0.5$
or we have optical photometry.  If we have color information, particularly a color
bracketing the $4000\AA$ spectral break, then the color information constrains
the redshift more strongly than the FP.  Figure 3 tests the accuracy of the method
by comparing the photometric redshift estimates to the known spectroscopic redshifts
for the lens and cluster galaxies. Keep in mind that these estimates are made with
1--3 filter photometry. 
 
For the lens galaxies, all the poor estimates with large redshift uncertainties are 
lenses at $z_l>0.5$ for which we have only infrared photometry.  With the addition of any optical data
the uncertainties collapse and the prediction usually matches the measured redshift.
In a few cases (Q0142--100 and B1608+656 in Fig. 3), we have accurate 3-filter photometry
and small redshift uncertainties, but the predicted redshift is incorrect.  For 
B1608+656, the problem is that the galaxy is an E+A galaxy with more recent star formation 
(Myers et al. 1995).
The average accuracies, $\langle z_{FP}-z_{true}\rangle$, are $-0.03\pm0.10$
and $-0.02\pm0.07$ for the lens and cluster galaxies respectively.  Much of the
scatter for the lens galaxies is due to the systems with only H-band data.

\begin{figure}[ht]
\centerline{\psfig{figure=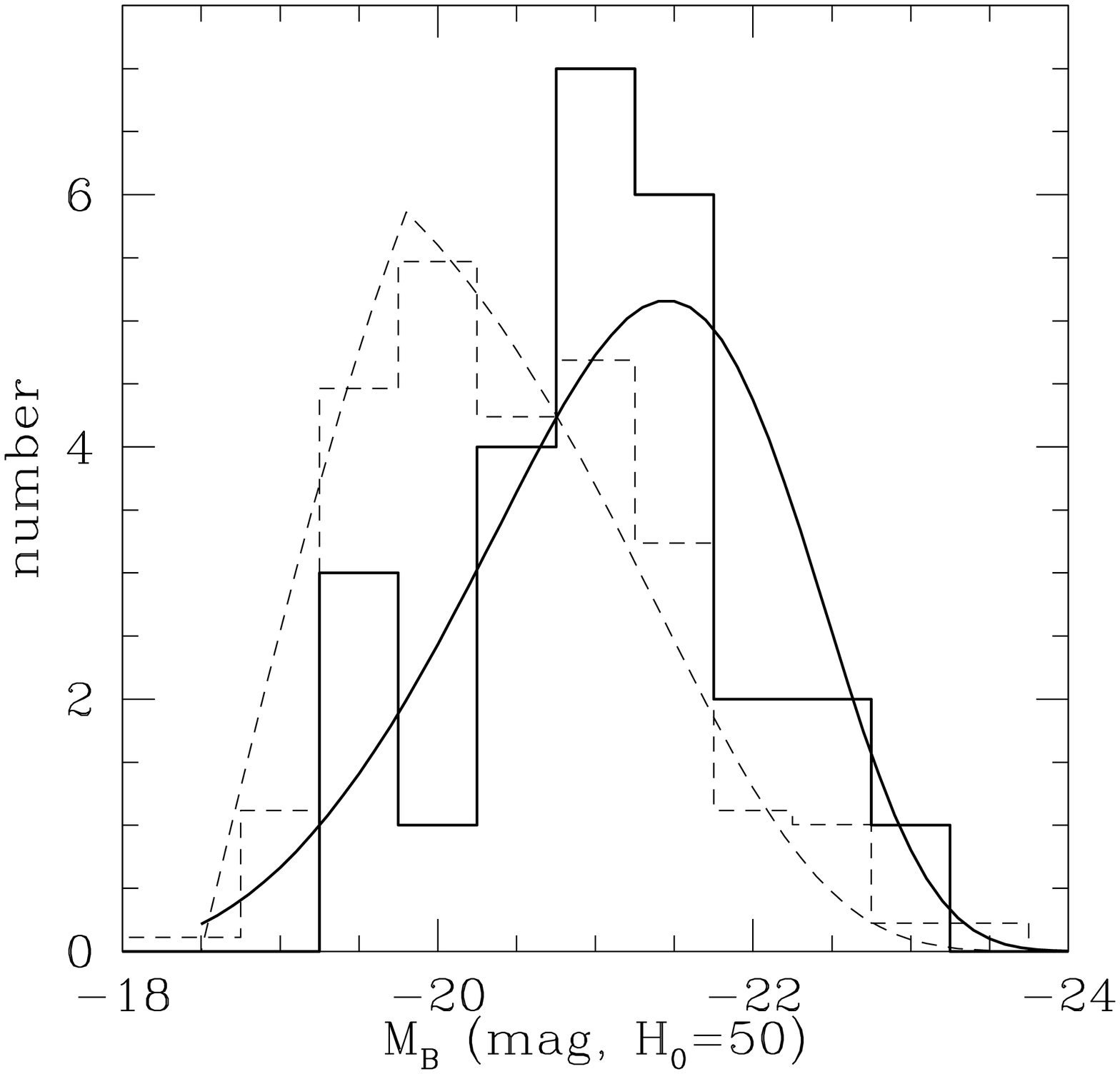,width=2.5in}
            \psfig{figure=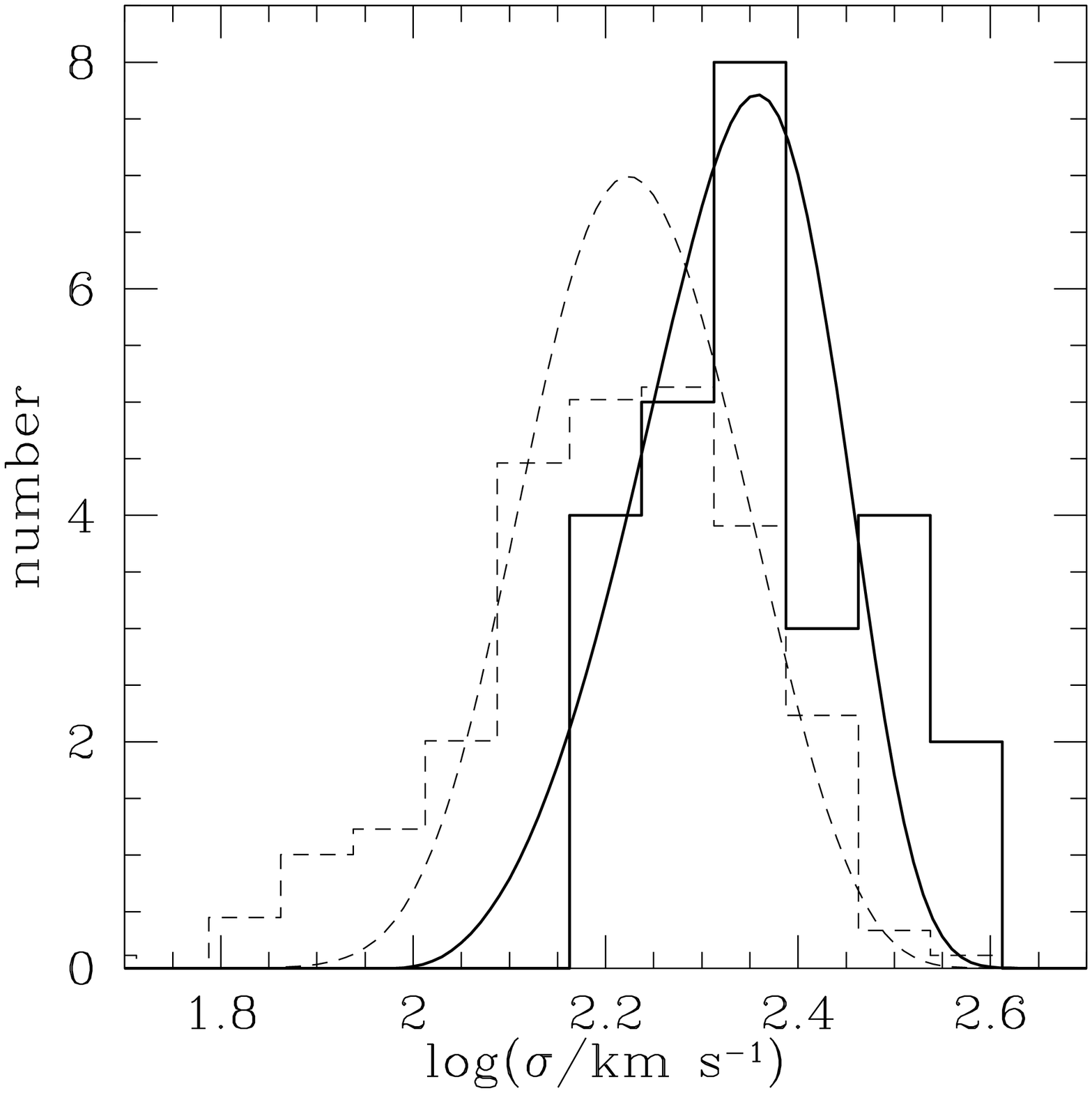,width=2.5in}}
\caption{The luminosity (left) and velocity (right) functions of lens galaxies
  compared to local clusters. The heavy solid (light dashed) histograms show the observed
  distributions of the lens (cluster) galaxies, and the heavy solid (light dashed) curves
  show the predicted distributions for the lens (cluster) galaxies. The sharp feature in
  the predicted luminosity function for the cluster galaxies is created by our {\it ad hoc}
  model for the magnitude limits in the JFK sample of cluster galaxies. }  
\end{figure}

\section{The Mass and Luminosity Function of Lens Galaxies}

The lens sample is large enough to begin studying the luminosity and mass functions
of the lens galaxies.  The average image separation of the lenses is already the most
accurate way of estimating the mean velocity dispersion of massive galaxies, which
makes matching the image separations a critical component of any estimate of the
cosmological model using gravitational lens statistics. Because the lenses are a
mass-weighted galaxy sample, they should be more luminous and more massive than a 
magnitude limited sample of cluster galaxies.  We can make a preliminary comparison by 
assuming the standard model we have used for gravitational lens statistics (Kochanek 1996,
Falco et al. 1998).

We start with a standard Schechter luminosity function to count the galaxies,
\begin{equation}
  d n/dL = (n_*/L_*)(L/L_*)^\alpha \exp(-L/L_*),
\end{equation}
combined with a Faber-Jackson law to relate luminosity and velocity dispersion,
\begin{equation}
  L/L_* = (\sigma/\sigma_*)^\gamma,
\end{equation}
where the velocity dispersion is related to the image separation through eqn. (5).
The multiple imaging cross section, $\propto \Delta\theta^2 \propto \sigma^4 \propto L^{4/\gamma}$,
is used to compute the probability that a galaxy acts as a lens.  We fix the parameters
to the standard model we have used to study lens statistics ($\gamma=4$, $\alpha=-1$, 
$\sigma_*=225$~km~s$^{-1}$, and $B_*=-19.9+5\log h$ mag).  The galaxy density $n_*$ will be 
nearly degenerate with the cosmological model, so we will not examine it here.  

Both the lens and cluster galaxy samples are affected by selection effects.  For the local
JFK cluster galaxy sample, it is the magnitude limit of the sample, which we have
treated only approximately.  For the lens galaxies, the only important selection effect 
is the finite angular resolution of the surveys, which we include as a minimum detectable 
separation $\Delta\theta_{min}$.  The minimum separation leads to a minimum lens luminosity
$L_{min}/L_*=(\Delta\theta_{min}/\Delta\theta_*)^{\gamma/2}$ and a selection function
$S(x=L/L_{min})=x^3(10-15x+6x^2)$.  Combining all these effects and using our standard
parameter values, the distribution of all early-type galaxies is 
\begin{equation}
   dn/dM \propto dn /d\log\sigma \propto \exp(-L/L_*) \qquad L > L_{min}
\end{equation}
where $L_{min}$ is set by the survey magnitude limit,
and the distribution of lens galaxies is
\begin{equation}
   dn/dM \propto dn /d\log\sigma \propto (L/L_*) \exp(-L/L_*) S(L/L_{min})
\end{equation}
where $L_{min}$ is set by the survey resolution.
For the local cluster sample, the distribution of galaxies peaks at the luminosity
$L_{min}$ corresponding to the sample magnitude limit.  
The lens galaxy sample peaks at $L \simeq L_*$, while the selection function
matters only near $L_{min} \simeq 0.05 L_*$.  In fact, our model for the selection
function of the lens galaxies is better than that for the JFK cluster galaxies.
The results are shown in Figure 4, where we present histograms of the observed lens and
JFK cluster galaxy distributions as compared to our simple model.  As expected, the
lens galaxies are both more luminous and more massive than typical early-type galaxies.
More importantly, the differences closely match the predictions of our standard model.

\section{Future Growth}

This is only an interim report on studies of the distribution and evolution of 
galaxies using gravitational lenses, since we have used only half the known 
lenses and the number of lenses is still growing rapidly.  Our progress is
limited by two major problems.  First, despite our best efforts, only about 
70\% of the lenses have archival or scheduled HST images good enough to do 
surface photometry of the 
lens galaxy even under a generous definition of ``good enough.''  Accurate 
measurements of galaxy evolution, mass-to-light ratios, luminosity and mass 
functions require higher precision data than the fractional-orbit snapshots 
commonly used to confirm lens candidates.  Host galaxy contamination, while 
scientifically valuable (see Bernstein et al. and Rix et al. in these 
proceedings), is also a serious problem for many radio lenses.   Second, the 
high level of redshift incompleteness is a major barrier to using 
gravitational lenses as astrophysical tools.  Even so, the sample of 30 
early-type galaxies we used to study galaxy evolution is 5 times larger than
the biggest published sample of field early-type galaxies with kinematic 
measurements at comparable redshifts.  We hope to solve the first problem
with the continuation of the CASTLES project into HST Cycle 9, and 
the missing redshift problem should soon be solved by the advent of so
many new large telescopes.

\medskip
\noindent{\bf Acknowledgements:} Support for the CASTLES project was
provided by NASA through grant numbers GO-7495 and GO-7887 from the
Space Telescope Science Institute which is operated by the Association
of Universities for Research in Astronomy, Inc. under NASA contract
NAS 5-26555.

\end{document}